\documentclass[journal,letterpaper,final,twocolumn]{IEEEtran} 

\usepackage[cmex10]{amsmath}
\usepackage{amssymb}

\usepackage{graphicx}
\usepackage{epstopdf}

\def\cA{\mathcal{A}}
\def\cB{\mathcal{B}}
\def\cX{\mathcal{X}}
\def\cY{\mathcal{Y}}
\def\cC{\mathcal{C}}
\def\F{\mathbb{F}}
\def\E{\mathbb{E}}

\newcommand{\defn}{\mathrel{\stackrel{\Delta}{=}}}
\newcommand{\sym}{\text{sym}}

\newcommand{\Qsym}{Q_\text{sym}}

\DeclareMathOperator{\argmax}{argmax}

\title{On the Origin of Polar Coding}
\author{\IEEEauthorblockN{Erdal Ar{\i}kan}\\
\IEEEauthorblockA{Department of Electrical and Electronics Engineering\\
	Bilkent University, Ankara, Turkey\\
Email: arikan@ee.bilkent.edu.tr}}

\begin{document}
\maketitle

\begin{abstract}
Polar coding was conceived originally as a technique for boosting the cutoff rate of sequential decoding, along the lines of earlier schemes of Pinsker and Massey. The key idea in boosting the cutoff rate is to take a vector channel (either given or artificially built),  split it into multiple correlated subchannels, and employ a separate sequential decoder on each subchannel. Polar coding was originally designed to be a low-complexity recursive channel combining and splitting operation of this type, to be used as the inner code in a concatenated scheme with outer convolutional coding and sequential decoding. However, the polar inner code turned out to be so effective that no outer code was actually needed to achieve the original aim of boosting the cutoff rate to channel capacity. This paper explains the cutoff rate considerations that motivated the development of polar coding.

\end{abstract}


\section{Introduction}\label{Sect:Introduction}

The most fundamental parameter regarding a communication channel is unquestionably its capacity $C$,
a concept introduced by Shannon \cite{Shannon1948} that marks the highest rate at which information can be transmitted reliably over the channel. 
Unfortunately, Shannon's methods that established capacity as an achievable limit were non-constructive in nature,
and the field of coding theory came into being with the agenda of turning Shannon's promise into practical reality.
Progress in coding theory was very rapid initially, with the first two decades producing some of the most innovative ideas in that field, but no truly practical capacity-achieving coding scheme emerged in this early period.
A satisfactory solution of the coding problem had to await the invention of turbo codes \cite{Berrou1996} in 1990s. 
Today, there are several classes of capacity-achieving codes, among them a refined version of Gallager's LDPC codes from 1960s \cite{Gallager1962}. 
(The fact that LDPC codes could approach capacity with feasible complexity was not realized until after their rediscovery in the mid-1990s.)
A story of coding theory from its inception until the attainment of the major goal of the field can be found in the excellent survey article \cite{CostelloForney2007}. 

A recent addition to the class of capacity-achieving coding techniques is polar coding \cite{ArikanIT2009}.
Polar coding was originally conceived as a method of boosting the channel cutoff rate $R_0$, a parameter that appears in two main roles in coding theory. 
First, in the context of random coding
and maximum likelihood (ML) decoding, $R_0$ governs the pairwise error probability $2^{-NR_0}$, which leads to the union bound 
\begin{align}\label{eq:UnionBound}
\overline{P}_e < 2^{-N(R_0-R)}
\end{align}
on the probability of ML decoding error $\overline{P}_e$ for a randomly selected code with $2^{NR}$ codewords. Second, in the context of sequential decoding, $R_0$ emerges as the ``computational cutoff rate'' beyond which sequential decoding---a decoding algorithm for convolutional codes---becomes computationally infeasible.

While $R_0$ has a fundamental character in its role as an error exponent as in \eqref{eq:UnionBound}, its significance as the cutoff rate of sequential decoding is a fragile one.
It has long been known that the cutoff rate of sequential decoding can be boosted by designing variants of sequential decoding that rely on various channel combining and splitting schemes to create correlated subchannels on which multiple sequential decoders are employed to achieve a sum cutoff rate that goes beyond the sum of the cutoff rates of the original memoryless channels used in the construction.
An early scheme of this type is due to Pinsker \cite{Pinsker}, who used a concatenated coding scheme with an inner block code and an outer sequential decoder to get arbitrarily high reliability at constant complexity per bit at any rate $R < C$; however, this scheme was not practical. Massey \cite{Massey81} subsequently described a scheme to boost the cutoff by splitting a nonbinary erasure channel into correlated binary erasure channels. 
We will discuss both of these schemes in detail, developing the insights that motivated the formulation of polar coding as a practical scheme for boosting the cutoff rate to its ultimate limit, the channel capacity $C$.

The account of polar coding given here is not intended to be the shortest or the most direct introduction to the subject. 
Rather, the goal is to give a historical account, highlighting the ideas that were essential in the course of developing polar codes, but have fallen aside as these codes took their final form.
On a personal note, my acquaintance with sequential decoding began in 1980s during my doctoral work \cite{arikan1985sequential} which was about sequential decoding for multiple access channels. Early on in this period, I became aware of the ``anomalous'' behavior of the cutoff rate, as exemplified in the papers by Pinsker and Massey cited above, and the resolution of the paradox surrounding the boosting of the cutoff rate has been a central theme of my research over the years. Polar coding is the end result of such efforts. 

The rest of this paper is organized as follows.
We discuss the role of $R_0$ in the context of ML decoding of block codes in Section~\ref{Sect:blockcoding} and its role in the context of sequential decoding of tree codes in Section~\ref{Sect:SD}.
In Section~\ref{Sect:boost}, we discuss the two methods by Pinsker and Massey mentioned above for boosting the cutoff rate of sequential decoding.
In Section~\ref{Sect:SCA}, we examine the successive-cancellation architecture as an alternative method for boosting the cutoff rate, and in Section~\ref{Sect:Polar} introduce polar coding 
as a special instance of that architecture.
The paper concludes in Section~\ref{Sect:Conclusion} with a summary.

Throughout the paper, we use the notation $W: \cX\to \cY$ to denote a discrete memoryless channel $W$ with input alphabet $\cX$, output alphabet $\cY$, and
channel transition probabilities $W(y|x)$ (the conditional probability that $y\in \cY$ is received given that $x\in \cX$ is transmitted).
We use the notation $a^N$ to denote a vector $(a_1,\ldots,a_N)$ and $a_{i}^j$ to denote a subvector $(a_i,\ldots,a_j)$.
All logarithms in the paper are to the base two.

\section{The parameter $R_0$}\label{Sect:blockcoding}

The goal of this section is to discuss the significance of the parameter $R_0$
in the context of block coding and ML decoding.
Throughout this section, let $W:\cX\to \cY$ be a fixed but arbitrary memoryless channel.
We will suppress in our notation the dependence of channel parameters on $W$ with the exception of some definitions that are referenced later in the paper.

\subsection{Random-coding bound}\label{Subsect:RandomCodingBound}
Consider a communication system using block coding at the transmitter and ML decoding at the receiver.
Specifically, suppose that the system employs a $(N,R)$ block code, {\sl i.e.}, a code of length $N$ and rate $R$, for transmitting one of $M\defn \lfloor 2^{NR}\rfloor $ messages by $N$ uses of $W$.
We denote such a code as a list of codewords $\cC=\{x^N(1),\ldots,x^N(M)\}$, where each codeword is an element of $\cX^N$.
Each use of this system comprises selecting, at the transmitter, a message $m\in \{1,\ldots,M\}$ at random from the uniform distribution, encoding it into the codeword $x^N(m)$, and sending $x^N(m)$ over $W$.
At the receiver, a channel output $y^N$ is observed with probability 
\begin{equation*}
W^N(y^N |x^N(m)) \defn \prod_{i=1}^N W(y_i|x_i(m))
\end{equation*}
and $y^N$ is mapped to a decision $\hat{m}$ by the ML rule
\begin{equation}\label{eq:ML}
\hat{m}(y^N) \defn \argmax_{m'} \,W^N(y^N|x^N(m')).
\end{equation}
The performance of the system is measured by the probability of ML decoding error, 
\begin{equation}\label{eq:PeC}
P_e (\cC) \defn \sum_{m=1}^M \frac{1}{M} \sum_{y^N: \hat{m}(y^N) \neq m} W^N(y^N|x^N(m)).
\end{equation}

Determining $P_e(\cC)$ for a specific code $\cC$ is a well-known intractable problem.
Shannon's {\sl random-coding} method \cite{Shannon1948} circumvents this difficulty by considering an ensemble of codes,
in which each individual code $\cC=\{x^N(1),\ldots,x^N(M)\}$ is regarded 
as a sample with a probability assignment
\begin{equation}\label{eq:RandCode}
\Pr(\cC) = \prod_{m=1}^M \prod_{n=1}^N Q(x_n(m),
\end{equation}
where $Q$ is a channel input distribution.
We will refer to this ensemble as a ``$(N,R,Q)$ ensemble''.
We will use the upper-case notation $X^N(m)$ to denote the random codeword for message $m$, viewing $x^N(m)$ as a realization of $X^N(m)$.
The product-form nature of the probability assignment \eqref{eq:RandCode} signifies that the array of $MN$ symbols $\{X_n(m); 1\le n\le N, 1\le m\le M\}$, constituting the code, are sampled in i.i.d. manner.

The probability of error averaged over the $(N,R,Q)$ ensemble is given by
\begin{equation*}
\overline{P}_e(N,R,Q) \,\defn \,\sum_{\cC} \Pr(\cC) P_e(\cC),
\end{equation*}
where the sum is over the set of all $(N,R)$ codes.
Classical results in information theory provide bounds of the form 
\begin{equation}\label{eq:RC}
\overline{P}_e(N,R,Q) \,\le\, 2^{-NE_r(R,Q)},
\end{equation}
where the function $E_r(R,Q)$ is a {\sl random-coding exponent},
whose exact form depends on the specific bounding method used.
For an early reference on such random-coding bounds, we refer to Fano \cite[Chapter~9]{FanoBook}, who also gives a historical account of the subject.
Here, we use a version of the random-coding bound due to Gallager \cite{gallager1965simple}, \cite[Theorem 5.6.5]{GallagerBook}),
in which the exponent is given by
\begin{equation}\label{eq:GallagerExponent}
E_r(R,Q) = \max_{0\le \rho \le 1} \left[E_0(\rho,Q) - \rho R\right],
\end{equation}
$$
E_0(\rho,Q) \defn -\log \sum_{y\in \cY} \bigg[\sum_{x\in \cX} Q(x) W(y|x)^{1/(1+\rho)}\bigg]^{1+\rho}.
$$
It is shown in \cite[pp.~141-142]{GallagerBook} that, for any fixed $Q$, $E_r(R,Q) >0 $ for all $R < C(Q)$, where $C(Q)=C(W,Q)$ is
the channel capacity with input distribution $Q$, defined as  
\begin{equation*}
C(W,Q) \; \defn \;\sum_{x,y} Q(x) W(y|x) \log \bigg(\frac{W(y|x)}{\sum_{x'} Q(x')W(y|x')}\bigg).
\end{equation*}
This establishes that reliable communication is possible for all rates $R < C(Q)$, with the probability of ML decoding error approaching zero exponentially in $N$.
Noting that the channel capacity is given by
\begin{equation}\label{eq:defC}
C = C(W) \defn \max_{Q} C(W,Q),
\end{equation}
the channel coding theorem follows as a corollary.

In a converse result \cite{Gallager73}, Gallager also shows that $E_r(R,Q)$ is the best possible exponent of its type in the sense that 
$$
-\frac1N \,\log \overline{P}_e(N,R,Q) \to E_r(R,Q)
$$ 
for any fixed $0\le R \le C(Q)$. This converse shows that $E_r(R,Q)$ is a channel parameter of fundamental significance in ML decoding of block codes.

For the best random-coding exponent of the type \eqref{eq:GallagerExponent} for a given $R$, one may maximize
$E_r(R,Q)$ over $Q$, and obtain the optimal exponent as
\begin{equation}\label{eq:randomcodingexponent}
E_r(R) \defn \max_{Q} E_r(R,Q).
\end{equation}

\begin{figure}[ht]
\begin{center}
\resizebox{!}{5cm}{
\includegraphics{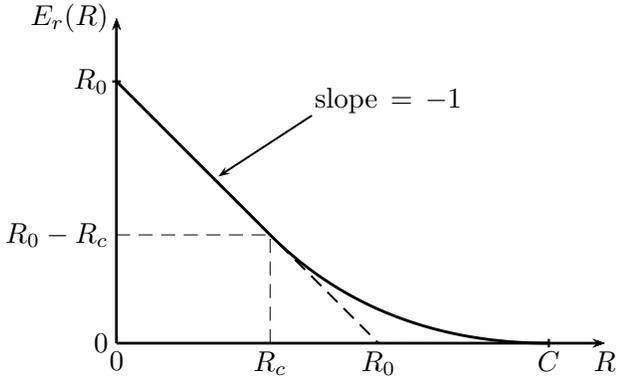}
}
\end{center}
\caption{Random-coding exponent $E_r(R)$ as a function of $R$ for a BSC.}
\label{fig:R0C}
\end{figure}

We conclude this part with an example.
Fig.~\ref{fig:R0C} gives a sketch of $E_r(R)$ for a binary symmetric channel (BSC), {\sl i.e.\/}, a channel $W:\cX\to \cY$ with $\cX=\cY = \{0,1\}$ and $W(1|0)=W(0|1)=p$ for some fixed $0\le p\le \frac12$. In this example, the $Q$ that achieves the maximum in \eqref{eq:randomcodingexponent} happens to be
the uniform distribution, $Q(0)=Q(1)=\frac12$ \cite[p.~146]{GallagerBook}, as one might expect due to symmetry. 

The figure shows that $E_r(R)$ is a convex function, starting at a maximum value $R_0\defn E_r(0)$ at $R=0$ and decreasing to 0 at $R=C$.
The exponent has a straight-line portion for a range of rates $0\le R\le R_c$, where the slope is $-1$.
The parameters $R_0$ and $R_c$ are called, respectively, the {\sl cutoff rate} and the {\sl critical rate}.
The union bound \eqref{eq:UnionBound} coincides with the random-coding bound over the straight-line portion of the latter,
becomes suboptimal in the range $R_c< R\le R_0$ (shown as a dashed line in the figure), and useless for $R>R_0$. 
These characteristics of $E_r(R)$ and its relation to $R_0$ are general properties that hold for all channels. 
In the rest of this section, we focus on $R_0$ and discuss it from various other aspects to gain a better understanding of this ubiquitous parameter.

\subsection{The union bound}
In general, the union bound is defined as
\begin{align}\label{eq:unionbound}
\overline{P}_e(N,R,Q) & < 2^{-N[R_0(Q)-R]},
\end{align}
where $R_0(Q) = R_0(W,Q)$ is the channel cutoff rate with input distribution $Q$, defined as
\begin{equation}\label{eq:defR0Q}
R_0(W,Q) \; \defn \; -\log \sum_{y\in \cY}\bigg[ \sum_{x\in \cX} Q(x)\sqrt{W(y|x)}\bigg]^2.
\end{equation}
The union bound \eqref{eq:unionbound} may be obtained from the random-coding bound by setting $\rho=1$ in \eqref{eq:GallagerExponent}, instead of maximizing over $0\le \rho\le 1$
and noticing that $R_0(Q)$ equals $E_0(1,Q)$.
The union bound and the random-coding bound coincide over a range of rates $0\le R\le R_c(Q)$, where $R_c(Q)$ is the critical rate at input distribution $Q$.
The tightest form of the union bound \eqref{eq:unionbound} is obtained by using an input distribution $Q$ that maximizes $R_0(Q)$,
in which case we obtain the usual form of the union bound as given by \eqref{eq:UnionBound} with 
\begin{equation}\label{eq:defR0}
R_0 = R_0(W) \defn \max_Q \; R_0(W,Q).
\end{equation}

The role of $R_0$ in connection with random-coding bounds can be understood by looking at the pairwise error probabilities under ML decoding.
To discuss this, consider a specific code $\cC=\{x^N(1),\ldots,x^N(M)\}$ and fix two distinct messages $m\neq m'$, $1\le m,m'\le M$.
Let $P_{m,m'}(\cC)$ denote the probability of pairwise error, namely, the probability that the erroneous message $m'$ appears to an ML decoder at least as likely as the correct message $m$;
more precisely,
\begin{equation}\label{eq:defPmm'}
P_{m,m'}(\cC) \; \defn \; \sum_{y^N\in E_{m,m'}(\cC)} W^N(y^N|x^N(m)),
\end{equation}
where $E_{m,m'}(\cC)$ is a pairwise error event defined as
$$
E_{m,m'}(\cC) \defn \big\{y^N: W^N(y^N|x^N(m')) \ge W^N(y^N|x^N(m))\big\}.
$$
Although $P_{m,m'}(\cC)$ is difficult to compute for specific codes,
its ensemble average,
\begin{equation}\label{eq:defaveragePmm'}
\overline{P}_{m,m'}(N,Q) \; \defn \; \sum_{\cC} \Pr(\cC) P_{m,m'}(\cC),
\end{equation}
is bounded as
\begin{equation}\label{eq:pairwiseerrorbound}
\overline{P}_{m,m'}(N,Q) \; \le \; 2^{-NR_0(Q)}, \quad m\neq m'.
\end{equation}
We provide a proof of inequality \eqref{eq:pairwiseerrorbound} in the Appendix to show that this well-known and basic result about the cutoff rate can be proved easily from first principles. 
 
The union bound \eqref{eq:unionbound} now follows from \eqref{eq:pairwiseerrorbound} by noting that an ML error occurs only if some pairwise error occurs:
\begin{align*}
\overline{P}_e(N,R,Q) & \; \le \; \sum_{m}\frac1M \sum_{m'\neq m} \overline{P}_{m,m'}(N,Q)\\
& \le \; (2^{NR}-1)\, 2^{-NR_0(Q)} < 2^{-N[R_0(Q)-R]}.
\end{align*}

This completes our discussion of the significance of $R_0$ as an error exponent in random-coding.
In summary, $R_0$ governs the random-coding exponent at low rates and is fundamental in that sense. 

For future reference, we note here that, when $W$ is a binary-input channel with $\cX=\{0,1\}$, the cutoff rate expression \eqref{eq:defR0Q} simplifies to
\begin{equation}\label{eq:defR0BinaryQ}
R_0(W,Q) = - \log (1 - q + qZ)
\end{equation}
where $q\defn 2\,Q(0)Q(1)$ and $Z$ is the channel {\sl Bhattacharyya parameter}, defined as
\begin{equation*}
Z = Z(W) \; \defn \; \sum_{y\in \cY} \sqrt{W(y|0)W(y|1)}.
\end{equation*}

\subsection{Guesswork and $R_0$}\label{Subsect:Guessing}
Consider a coding system identical to the one described in the preceding subsection, except now suppose that the decoder is a {\sl guessing} decoder.
Given the channel output $y^N$, a guessing decoder is allowed to produce a sequence of guesses $m_1(y^N),m_2(y^N),\ldots$ at the correct message $m$ until a helpful ``genie'' tells the decoder when to stop. 
More specifically, after the first guess $m_1(y^N)$ is submitted, the genie tells the decoder if $m_1(y^N)=m$; if so, decoding is completed; otherwise, the second guess $m_2(y^N)$ is submitted; and, so on. The operation continues until the decoder produces the correct message. We assume that the decoder never repeats an earlier guess,
so the task is completed after at most $M$ guesses. 

An appropriate ``score'' for such a guessing decoder is the {\sl guesswork}, which we define as the number of {\sl incorrect} guesses $G_0(\cC)$ until completion of the decoding task. 
The guesswork $G_0(\cC)$ is a random variable taking values in the range $0$ to $M-1$.
It should be clear that the optimal strategy for minimizing the average guesswork is to use the ML order: namely, to set the first guess $m_1(y^N)$ equal to a most likely message given $y^N$ (as in \eqref{eq:ML}), the second guess $m_2(y^N)$ equal to a second most likely message given $y^N$, {\sl etc}.
We call a guessing decoder of this type an {\sl ML-type} guessing decoder.

Let $\E[G_0(\cC)]$ denote the average guesswork for an ML-type guessing decoder for a specific code $\cC$.
We observe that an incorrect message $m'\neq m$ precedes the correct message $m$ in the ML guessing order only if a channel output $y^N$ is received such that
$W^N(y^N|x^N(m'))\ge W^N(y^N|x^N(m))$; thus, $m'$ contributes to the guesswork only if a pairwise error event takes place between the correct message $m$ and the incorrect message $m'$.
Hence, we have
\begin{equation}\label{eq:Guesswork}
\E[G_0(\cC)] = \sum_{m}\frac1M \sum_{m'\neq m} P_{m,m'}(\cC)
\end{equation}
where $P_{m,m'}(\cC)$ is the pairwise error probability under ML decoding as defined in \eqref{eq:defPmm'}.
We observe that the right side of \eqref{eq:Guesswork} is the same as the union bound on the probability of ML decoding error for code $\cC$.
As in the union bound, rather than trying to compute the guesswork for a specific code, we consider
the ensemble average over all codes in a $(N,R,Q)$ code ensemble,
and obtain
\begin{align*}
\overline{G}_0(N,R,Q)) & \; \defn \; \sum_{\cC}\Pr(\cC)\E[G_0(\cC)] \\
& = \; \sum_{m}\frac1M \sum_{m'\neq m} \overline{P}_{m,m'}(N,Q),
\end{align*}
which in turn simplifies by \eqref{eq:pairwiseerrorbound} to
\begin{equation*}
\overline{G}_0(N,R,Q) \; \le  \; 2^{N[R-R_0(Q)]}.
\end{equation*}
The bound on the guesswork is minimized if we use an ensemble $(N,R,Q^*)$ for which $Q^*$ achieves the maximum of $R_0(Q)$ over all $Q$; in that case, 
the bound becomes
\begin{equation}\label{eq:guessworkUB}
\overline{G}_0(N,R,Q^*)\; \le  \; 2^{N[R-R_0]}.
\end{equation}

In \cite{Arikan96}, the following converse result was provided for {\sl any} code $\cC$ of rate $R$ and block length $N$:
\begin{equation}\label{eq:guessworkLB}
\E[G_0(\cC)] \ge \max\{0,2^{N(R-R_0-o(N))}-1\},
\end{equation}
where $o(N)$ is a quantity that goes to 0 as $N$ becomes large.
Viewed together, \eqref{eq:guessworkUB} and \eqref{eq:guessworkLB} state that $R_0$ is a rate threshold that separates two very distinct regimes of operation for
an ML-type guessing decoder on channel $W$:
for $R> R_0$, the average guesswork is exponentially large in $N$ regardless of how the code is chosen; for $R <R_0$, it is possible to keep the average guesswork close to 0
by an appropriate choice of the code.
In this sense, $R_0$ acts as a {\sl computational cutoff rate}, beyond which guessing decoders become computationally infeasible.

Although a guessing decoder with a genie is an artificial construct, it provides a valid computational model for studying the computational complexity of the sequential
 decoding algorithm, as we will see in Sect.~\ref{Sect:SD}. The interpretation of $R_0$ as a computational cutoff rate in guessing will carry over directly to sequential decoding.

\section{Sequential decoding}\label{Sect:SD}

The random-coding results show that a code picked at random is likely to be a very good code with an ML decoder error probability exponentially small in code block length.
Unfortunately, randomly-chosen codes do not solve the coding problem, because such codes lack structure, which makes them hard to encode and decode.
For a practically acceptable balance between performance and complexity, there are two broad classes of techniques. One is the algebraic-coding approach that eliminates random elements from code construction entirely; this approach has produced many codes with low-complexity encoding and decoding algorithms, but so far none that is capacity-achieving
with a practical decoding algorithm. The second is the probabilistic-coding approach that retains a certain amount of randomness while imposing a significant degree of structure on the code so that low-complexity encoding and decoding are possible.
A tree code with sequential decoding is an example of this second approach.

\subsection{Tree codes}
A tree code is a code in which the codewords conform to a tree structure.
A convolutional code is a tree code in which the codewords are closed under vector addition. These codes were introduced by Elias \cite{Elias1955} with the motivation to reduce the complexity of ML decoding by imposing a tree structure on block codes.
In the discussion below, we will be considering tree codes with infinite length and infinite memory in order to avoid distracting details; although, in practice, one would use a finite-length finite-memory convolutional code.

\begin{figure}[ht]
\begin{center}
\resizebox{!}{10cm}{
\includegraphics{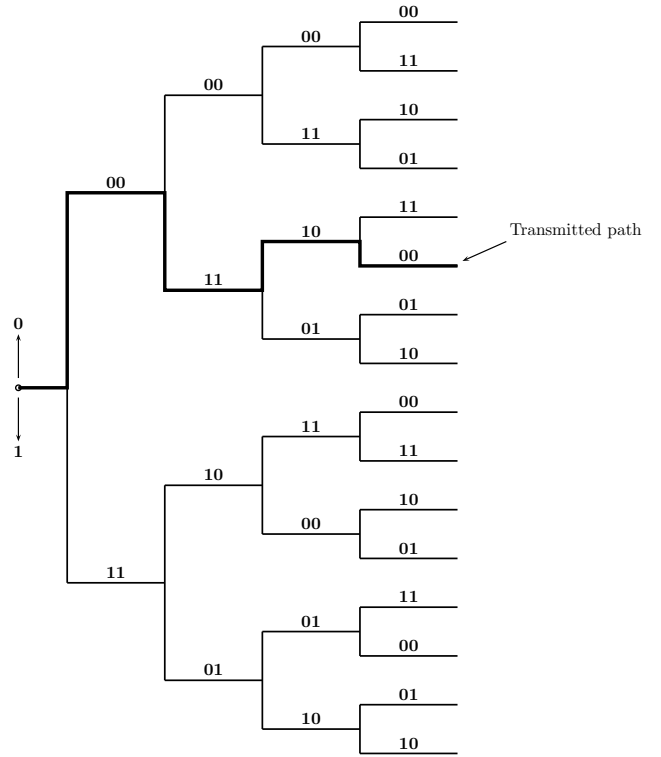}
}
\end{center}
\caption{A tree code.}
\label{fig:TreeCode}
\end{figure}

The encoding operation for tree codes can be described with the aid of Fig.~\ref{fig:TreeCode},
which shows the first four levels of a tree code with rate $R=1/2$.
Initially, the encoder is situated at the root of the tree and the codeword string is empty.
During each unit of time, one new data bit enters the encoder and causes it to move one level deeper into the tree,
taking the upper branch if the input bit is 0, the lower one otherwise. 
As the encoder moves from one level to the next, it puts out the two-bit label on the traversed branch as the current segment of the codeword. 
An example of such encoding is shown in the figure, where in response to the input string 0101 the encoder produces 00111000.

The decoding task for a tree code may be regarded as a search for the correct (transmitted) path through the code tree, given a noisy observation of that path. 
Elias \cite{Elias1955} gave a random-coding argument showing that tree codes are capacity-achieving. (In fact, he proved this result also for time-varying convolutional codes.)
Thus, having a tree structure in a code comes with no penalty in terms of capacity, but makes it possible to implement ML decoding at reduced complexity thanks to various
search heuristics that exploit this structure.

\subsection{Sequential decoding}

Consider implementing ML decoding in an infinite tree code. 
Clearly, one cannot wait until the end of transmissions, decoding has to start with a partial (and noisy) observation
of the transmitted path. 
Accordingly, it is reasonable to look for a decoder that has, at each time instant, a working hypothesis with respect to the transmitted path but is permitted to go back and change that hypothesis as new observations arrive from the channel. There is no final decision in this framework; all hypotheses are tentative.

An algorithm of this type, called {\sl sequential decoding}, was introduced by Wozencraft \cite{Wozencraft1957}, \cite{Wozencraft1961},
and remained an important research area for more than a decade.
A version of sequential decoding due to Fano \cite{Fano1963} was used in the Pioneer 9 deep-space mission in the late 1960s \cite{layland1971flexible}.
Following this brief period of popularity, sequential decoding was eclipsed by other methods and never recovered. 
(For a perspective on the rise and fall of sequential decoding, we refer to \cite{Forney1995}.)

The main drawback of sequential decoding, which partly explains its decline, is the variability of computation. 
Sequential decoding is an ML algorithm, capable of producing error-free output given enough time, 
but this performance comes at the expense of using a backtracking search. The time lost in backtracking increases with the severity of noise in the channel and the rate of the code. 
From the very beginning, it was recognized \cite{Wozencraft1957}, \cite{Reiffen1960} that the computational complexity of sequential decoding is characterized by the existence of a computational cutoff rate, denoted $R_{\text{cutoff}}$ (or $R_{\text{comp}}$), that separates two radically different regimes of operation in terms of complexity: at rates $R < R_{\text{cutoff}}$ the average number of decoding operations per bit remains bounded by a constant, while for $R > R_{\text{cutoff}}$, the decoding latency grows arbitrarily large. 
Later work on sequential decoding established that $R_{\text{cutoff}}$ coincides with the channel parameter $R_0$.
For a proof of the achievability part, $R_{\text{cutoff}} \ge R_0$, and bibliographic references, we refer to \cite[Theorem 6.9.2]{GallagerBook}; for the converse, 
$R_{\text{cutoff}}\le R_0$, we refer to \cite{JacobsBerlekamp}, \cite{Arikan1988}.

\begin{figure}[ht]
\begin{center}
\resizebox{!}{2in}{
\includegraphics{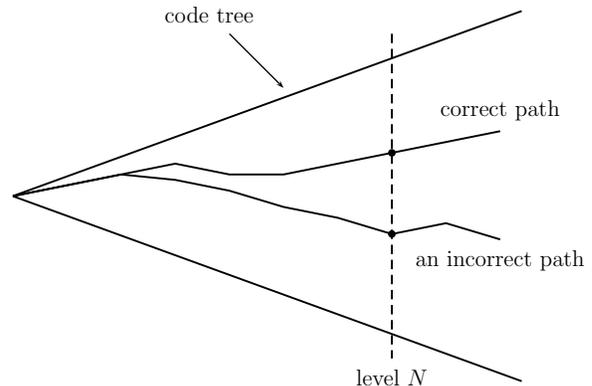}
}
\end{center}
\caption{Searching for the correct node at level $N$.}
\label{fig:TreeCodeSearch}
\end{figure}

An argument that explains why ``$R_{\text{cutoff}} = R_0$'' can be given by a simplified complexity model introduced by Jacobs and Berlekamp \cite{JacobsBerlekamp} that abstracts out the essential features of sequential decoding while leaving out irrelevant details.
In this simplified model one fixes an arbitrary level $N$ in the code tree, as in Fig.~\ref{fig:TreeCodeSearch},
and watches decoder actions only at this level.
The decoder visits level $N$ a number of times over the span of decoding, paying various numbers of visits to various nodes.
A sequential decoder restricted to its operations at level $N$ may be seen as a type of guessing decoder in the sense of Sect.~\ref{Subsect:Guessing}
operating on the block code of length $N$ obtained by truncating the tree code at level $N$.
Unlike the guessing decoder for a block code, a sequential decoder does not need a genie to find out whether its current guess is correct; an incorrect turn by the sequential decoder is sooner or later detected with probability one with the aid of a {\sl metric}, {\sl i.e.}, a likelihood measure that tends to decrease as soon as the decoder deviates from the correct path.
To follow the guessing decoder analogy further, let $G_{0,N}$ be the number of distinct nodes visited at level $N$ by the sequential decoder before its first visit to the correct node at that level.
In light of the results given earlier for the guessing decoder, it should not be surprising that $\E[G_{0,N}]$ shows two types of behavior depending on the rate: for $R> R_0$, $\E[G_{0,N}]$ grows exponentially with $N$; for $R < R_0$, $\E[G_{0,N}]$ remains bounded by a constant independent of $N$.
Thus, it is natural that $R_{\text{cutoff}} = R_0$.

To summarize, this section has explained why $R_0$ appears as a cutoff rate in sequential decoding by linking sequential decoding to guessing.
While $R_0$ has a firm meaning in its role as part of the random-coding exponent, it is a fragile parameter as the cutoff rate in sequential decoding.
By devising variants of sequential decoding, it is possible to break the $R_0$ barrier, as examples in the next section will demonstrate.

\section{Boosting the cutoff rate in sequential decoding}\label{Sect:boost}

In this section, we discuss two methods for boosting the cutoff rate in sequential decoding.
The first method, due to Pinsker \cite{Pinsker}, was introduced in the context of a theoretical analysis of the tradeoff between complexity and performance in coding.   
The second method, due to Massey \cite{Massey81}, had more immediate practical goals and was introduced in the context of the design of a coding and modulation scheme for an optical channel.
We present these schemes in reverse chronological order since Massey's scheme is simpler and contains the prototypical idea for boosting the cutoff rate.

\subsection{Massey's scheme}\label{Sect:Massey}

\begin{figure}[ht!]
\begin{center}
\hspace*{0cm}\resizebox{!}{1.50in}{
\includegraphics{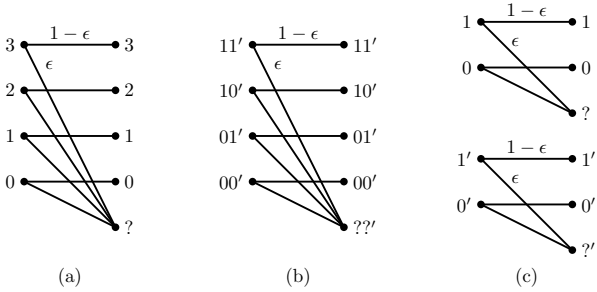}
}
\caption{Splitting a QEC into two fully-correlated BECs by input-relabeling.}
\label{fig:Massey1}
\end{center}
\end{figure}

A paper by Massey \cite{Massey81} revealed a truly interesting aspect of the 
cutoff rate by showing that it could be boosted by simply ``splitting'' a given channel.
The simplest channel where Massey's idea can be employed is a 
quaternary erasure channel (QEC) with erasure probability $\epsilon$, as shown in Fig.~\ref{fig:Massey1}(a). 
The capacity and cutoff rate of this channel are given by
$C_{\text{QEC}}(\epsilon) = 2(1-\epsilon)$ and $R_{0,\text{QEC}}(\epsilon) =\log \frac{4}{1+3\epsilon}$.

Consider relabeling the inputs of the QEC with a pair of bits as in Fig.~\ref{fig:Massey1}(b).
This turns the QEC into a vector channel with input $(b,b')$ and output $(s,s')$
and transition probabilities
\begin{align*}
(s,s') & = 
\begin{cases} 
(b,b'), & \text{with probability $1-\epsilon$},\\
(?,?'), & \text{with probability $\epsilon$}.
\end{cases}
\end{align*}
Following such relabeling, we can split the QEC into two binary erasure channels (BECs), as shown in Fig.~\ref{fig:Massey1}(c).
The resulting BECs are fully correlated in the sense that an erasure occurs in one if and only if an erasure occurs in the other.

\begin{figure}[ht!]
\begin{center}
\resizebox{!}{1.5in}{
\includegraphics{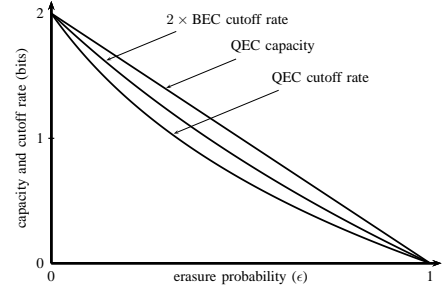}
}
\end{center}
\caption{Capacity and cutoff rates with and without splitting a QEC.}
\label{fig:Massey3}
\end{figure}

One way to employ coding on the original QEC is to split it as above into two BECs and employ coding on each BEC independently, ignoring the correlation between them.
In that case, the achievable sum capacity is given by $2C_{\text{BEC}}(\epsilon) = 2(1-\epsilon)$, which is the same as the capacity of the original QEC.
Even more surprisingly, the achievable sum cutoff rate after splitting is $2R_{0,\text{BEC}}(\epsilon) = 2\log \frac{2}{1+\epsilon}$, 
which is strictly larger than $R_{0,\text{QEC}}(\epsilon)$ for any $0<\epsilon<1$. 
The capacity and cutoff rates for the two coding alternatives are sketched in Fig.~\ref{fig:Massey3},
showing that substantial gains in the cutoff rate are obtained by splitting.

The above example demonstrates in very simple terms that just by splitting a composite channel into its constituent subchannels one may be able 
obtain a net gain in the cutoff rate without sacrificing capacity. 
Unfortunately, it is not clear how to generalize Massey's idea to other channels.
For example, if the channel has a binary input alphabet, it cannot be split.
Even if the original channel is amenable to splitting, ignoring the correlations among the subchannels created by splitting may be costly in terms of capacity.
So, Massey's scheme remains an interesting isolated instance of cutoff-rate boosting by channel splitting.
Its main value lies in its simplicity and the suggestion that building correlated subchannels may be the key to achieving cutoff rate gains.
In closing, we refer to \cite{gallager1985perspective} for an alternative discussion of Massey's scheme from the viewpoint of multi-access channels.

\subsection{Pinsker's method}\label{Sect:Pinsker}
\begin{figure*}[!ht]
\begin{center}
\hspace*{-0.4cm}\resizebox{!}{2.5in}{
\includegraphics{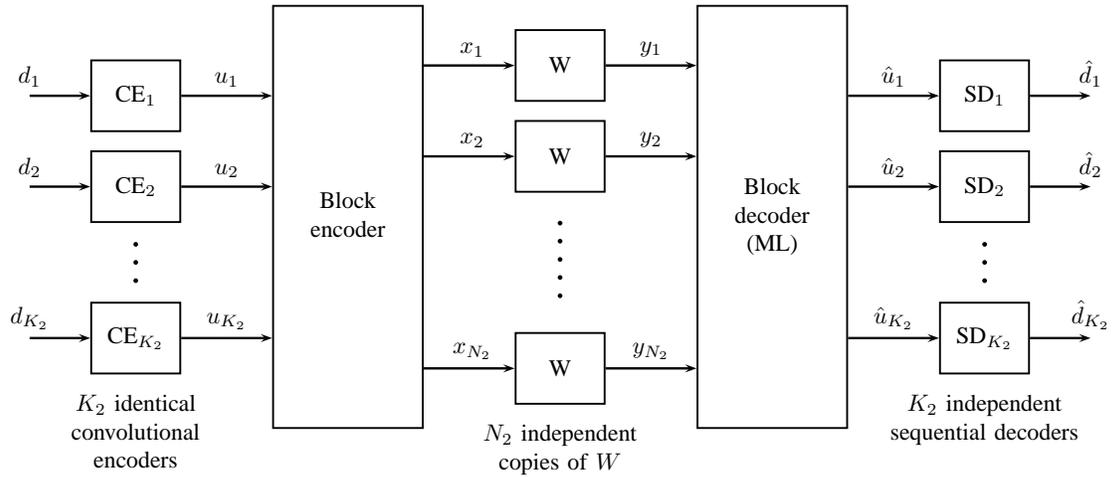}
}
\end{center}
\caption{Pinsker's scheme for boosting the cutoff rate.}
\label{Fig:Pinsker}
\end{figure*}

Pinsker was perhaps the first to draw attention to the flaky nature of the cutoff rate and suggest a general method to turn that into an advantage in terms of complexity of decoding.
Pinsker's scheme, shown in Fig.~\ref{Fig:Pinsker}, combines sequential decoding with Elias' product-coding method \cite{Elias1954}.
The main idea in Pinsker's scheme is to have an inner block code clean up the channels seen by a bank of outer sequential decoders, boosting the cutoff rate seen by each sequential decoder to near 1 bit.
In turn, the sequential decoders boost the reliability to arbitrarily high levels at low complexity.
Stated roughly, Pinsker showed that his scheme can operate arbitrarily close to capacity while providing arbitrarily low probability of error at constant average complexity per decoded bit.
The details are as follows.

Following Pinsker's exposition, we will assume that the channel $W$ in the system is a BSC with crossover probability $0\le p\le 1$,
in which case the capacity is given by 
$$
C(p)\defn 1 + p\log p + (1-p) \log(1-p).
$$ 
The user data consists of $K_2$ independent bit-streams, denoted $d_1,d_2,\ldots,d_{K_2}$.
Each stream is encoded by a separate convolutional encoder (CE), with all CEs operating at a common rate $R_1$. 
Each block of $K_2$ bits coming out of the CEs is encoded by an inner block code, which operates at rate $R_2= K_2/N_2$ and is assumed to be a linear code.
Thus, the overall transmission rate is $R = R_1\,R_2$.

The codewords of the inner block code are sent over $W$ by $N_2$ uses of that channel as shown in the figure. The received sequence is first passed through an ML decoder for the inner block code, 
then each bit obtained at the output of the ML decoder is fed into a separate sequential decoder (SD),
with the $i$th SD generating an estimate $\hat{d}_i$ of $d_i$, $1\le i\le K_2$. The SDs operate in parallel and independently (without exchanging any information). 
An error is said to occur if $\hat{d}_i\neq d_i$ for some $1\le i\le K_2$.

The probability of ML decoding error for the inner code, $p_e\defn P(\hat{u}^N\neq u^N)$, is independent of the transmitted codeword since the code is {\sl linear} and the channel is a BSC\footnote{The important point here is that the channel be {\sl symmetric} in the sense defined later in Sect.\ref{Sect:SCA}. Pinsker's arguments hold for any binary-input channel that is symmetric.}. 
Each frame error in the inner code causes a burst of bit errors that spread across the $K_2$ parallel bit-channels, but do not affect more than one bit in each channel
thanks to interleaving of bits by the product code.
Thus, each bit-channel is a memoryless BSC with a certain crossover probability, $p_i$, that equals the bit-error rate $P(\hat{u}_i\neq u_i)$ on the $i$th coordinate of the inner block code.
So, the the cutoff rate ``seen'' by the $i$th CE-SD pair is 
$$
R_0(p_i) \defn 1 -\log(1 +2\sqrt{p_i(1-p_i)}),
$$
which is obtained from \eqref{eq:defR0BinaryQ} with $Q(0)=Q(1)=\frac12$.
These cutoff rates are uniformly good in the sense that
$$
R_0(p_i) \ge R_0(p_e), \quad 1\le i\le K_2,
$$
since $0\le p_i\le p_e$, as in any block code.

It follows that the aggregate cutoff rate of the outer bit-channels is $\ge K_2R_0(p_e)$, which corresponds to a normalized cutoff rate of better than $R_2R_0(p_e)$ bits per channel use. 
Now, consider fixing $R_2$ just below capacity $C(p)$ and selecting $N_2$ large enough to ensure that $p_e\approx 0$.
Then, the normalized cutoff rate satisfies $R_2R_0(p_e)\approx C(p)$. 
This is the sense in which Pinsker's scheme boosts the cutoff rate to near capacity.

Although Pinsker's scheme shows that arbitrarily reliable communication at any rate below capacity is possible within constant complexity per bit, the ``constant'' entails the ML decoding complexity of an 
inner block code operating near capacity and providing near error-free communications. 
So, Pinsker's idea does not solve the coding problem in any practical sense.
However, it points in the right direction, suggesting that channel combining and splitting are the key to boosting the cutoff rate.

\subsection{Discussion}\label{Subsect:Conservation}

In this part, we discuss the two examples above in a more abstract way in order to identify the essential features that are behind the boosting of the cutoff rate.

\begin{figure}[ht]
\begin{center}
\resizebox{!}{1in}{
\includegraphics{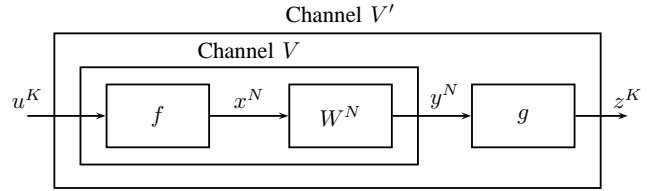}
}
\end{center}
\caption{Channels derived from $W$ by pre- and post-processing operations.}
\label{fig:DerivedChannel}
\end{figure}

Consider the system shown in Fig.~\ref{fig:DerivedChannel} that presents a framework general enough to accommodate both Pinsker's method and Massey's method as special instances.
The system consists of a {\sl mapper} $f$ and a {\sl demapper} $g$ that implement, respectively, the combining and splitting operations
for a given memoryless channel  $W:\cX\to \cY$.
The mapper and the demapper can be any functions of the form $f:\cA^K\to \cX^N$ and $g:\cY^N\to\cB^{K}$,
where the alphabets $\cA$ and $\cB$, as well as the dimensions $K$ and $N$ are design parameters.

The mapper acts as a pre-processor to create a derived channel $V:\cA^K \to \cY^{N}$ from vectors of length $K$ over $\cA$ to vectors of length $N$ over $\cY$.
The demapper acts as a post-processor to create from $V$ a second derived channel $V':\cA^K \to \cB^{K}$.
The well-known data-processing theorem of information theory \cite[p.~50]{GallagerBook} states that 
\begin{equation*}
C(V') \le C(V) \le N C(W),
\end{equation*}
where $C(V')$, $C(V)$, and $C(W)$ denote the capacities of $V'$, $V$, and $W$, respectively.
There is also a data-processing theorem that applies to the cutoff rate, stating that
\begin{equation}\label{eq:DP}
R_0(V') \le R_0(V)\le N R_0(W).
\end{equation}
The data-processing result for the cutoff rate follows from a more general result given in \cite[pp.~149-150]{GallagerBook} in the context of ``parallel channels.''
In words, inequality \eqref{eq:DP} states that it is impossible to boost the cutoff rate of a channel $W$ if one employs a {\sl single} sequential decoder on the channels $V$ or $V'$ derived from $W$ by any kind of pre-processing and post-processing operations.

On the other hand, there are cases where it is possible to split the derived channel $V'$ into $K$ memoryless channels
$V'_i:\cA\to \cB$, $1\le i\le K$, so that the normalized cutoff rate after splitting shows a cutoff rate gain in the sense that
\begin{equation}\label{eq:SubadditiveR0}
\frac1N\;\sum_{i=1}^K R_0(V'_i) > R_0(W).
\end{equation}
Both Pinsker's scheme and Massey's scheme are examples where \eqref{eq:SubadditiveR0} is satisfied.
In Pinsker's scheme, the alphabets are $\cA=\cB=\{0,1\}$, $f$ is an encoder for a binary block code of rate $K_2/N_2$, $g$ is an ML decoder for the block code, 
and the bit channel $V'_i$ is the channel between $u_i$ and $z_i$, $1\le i\le K_2$.
In Massey's scheme, with the QEC labeled as in Fig.~\ref{fig:Massey1}-(a), the length parameters are $K=2$ and $N=1$, $f$ is the identity map on $\cA=\{0,1\}^2$, and $g$ is the identity map on $\cB= \{0,1,?\}^2$.

As we conclude this section, a word of caution is necessary about the application of the above framework for cutoff rate gains.
The coordinate channels $\{V'_i\}$ created by the above scheme are in general not memoryless; they interfere with each other in complex ways depending on the specific $f$ and $g$ employed. 
For a channel $V_i'$ with memory, the parameter $R_0(V'_i)$ loses its operational meaning as the cutoff rate of sequential decoding.
Pinsker avoids such technical difficulties in his construction by using a {\sl linear} code and restricting the discussion to a {\sl symmetric} channel.
In designing systems that target cutoff rate gains as promised by \eqref{eq:SubadditiveR0}, these points should not be overlooked.

\section{Successive-cancellation architecture}\label{Sect:SCA}

In this section, we examine the successive-cancellation (SC) architecture, shown in Fig.~\ref{fig:DerivedChannel2}, as a general framework for boosting the cutoff rate.
The SC architecture is more flexible than Pinsker's architecture in Fig.~\ref{fig:DerivedChannel}, and may be regarded as a generalization of it.
This greater flexibility provides significant advantages in terms of building practical coding schemes, as we will see in the rest of the paper.
As usual, we will assume that the channel in the system is a binary-input channel $W:\cX=\{0,1\}\to \cY$.

\subsection{Channel combining and splitting}\label{Subsect:SCA}
\begin{figure*}[!t]
\begin{center}
\hspace*{0cm}\resizebox{!}{2.5in}{
\includegraphics{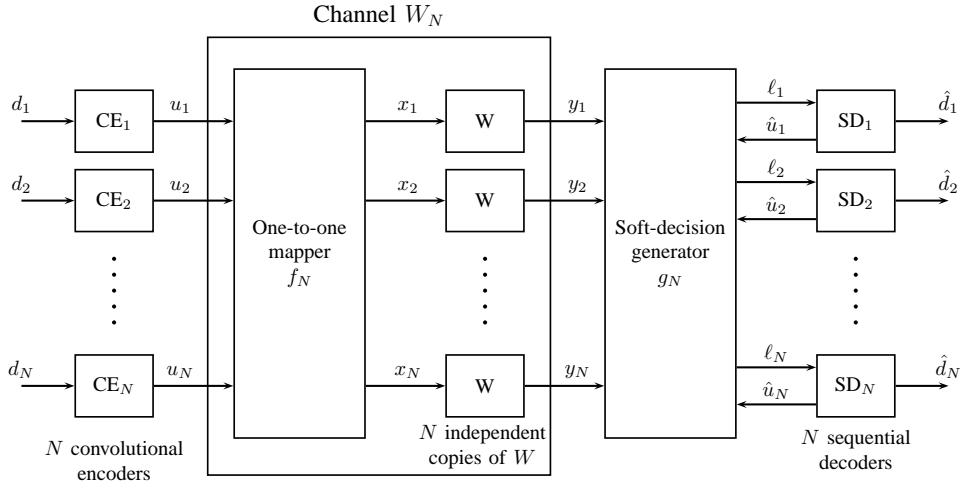}
}
\end{center}
\caption{Successive-cancellation architecture for boosting the cutoff rate.}
\label{fig:DerivedChannel2}
\end{figure*}

As seen in Fig.~\ref{fig:DerivedChannel2}, the transmitter in the SC architecture uses a 1-1 mapper $f_N$ that combines $N$ independent copies of $W$ to synthesize a channel 
$$
W_N: u^N\in \{0,1\}^N \to y^N\in \cY^N
$$ 
with transition probabilities
$$
W_N(y^N|u^N) = \prod_{i=1}^N W(y_i|x_i), \quad x^N =f_N(u^N).
$$

The SC architecture has room for $N$ CEs but these encoders do not have to operate at the same rate, which is one difference between the SC architecture and Pinsker's scheme.
The intended mode of operation in the SC architecture is to set the rate of the $i$th encoder CE$_i$ to a value commensurate with the capability of that channel.

The receiver side in the SC architecture consists of a soft-decision generator (SDG) and a chain of SDs that carry out SC decoding.
To discuss the details of the receiver operation, let us index the blocks in the system by $t$.
At time $t$, the $t$th code block $x^N$, denoted $x^N(t)$, is transmitted (over $N$ copies of $W$) and $y^N(t)$ is delivered to the receiver.
Assume that each round of transmission lasts for $T$ time units, with $\{x^N(1),\ldots,x^N(T)\}$ being sent and $\{y^N(1),\ldots,y^N(T)\}$ received.
Let us write $\{x^N(t)\}$ to denote $\{x^N(1),\ldots,x^N(T)\}$ briefly. Let us use similar time-indexing for all other signals in the system, for example,
let $d_i(t)$ denote the data at the input of the $i$th encoder CE$_{i}$ at time $t$.

Decoding in the SC architecture is done layer-by-layer, in $N$ layers: first, the data sequence $\{d_1(t):1\le t\le T\}$ is decoded, then $\{d_2(t)\}$ is decoded, and so on.
To decode the first layer of data $\{d_1(t)\}$, the SDG computes the soft-decision variables $\{\ell_1(t)\}$ as a function of $\{y^N(t)\}$ and feeds them into the first sequential decoder SD$_{1}$.
Given $\{\ell_1(t)\}$, SD$_1$ calculates two sequences: the estimates $\{\hat{d}_1(t)\}$ of $\{d_1(t)\}$, which it sends out as its final decisions about $\{d_1(t)\}$; 
and, the estimates $\{\hat{u}_1(t)\}$ of $\{u_1(t)\}$, which it feeds back to the SDG.
Having received $\{\hat{u}_1(t)\}$, the SDG proceeds to compute the soft-decision sequence $\{\ell_2(t)\}$ and feeds them into SD$_2$, which, in turn, computes the estimates 
$\{\hat{d}_2(t)\}$ and $\{\hat{u}_2(t)\}$, sends out $\{\hat{d}_2(t)\}$, and feeds $\{\hat{u}_2(t)\}$ back into the SDG.
In general, at the $i$th layer of SC decoding, the SDG computes the sequence $\{\ell_i(t)\}$ and feeds it to SD$_i$, which in turn computes a data decision sequence $\{\hat{d}_i(t)\}$, which it sends out, 
and a second decision sequence $\{\hat{u}_i(t)\}$, which it feeds back to SDG. The operation is completed when the $N$th decoder SD$_N$ computes and sends out the data decisions $\{\hat{d}_N(t)\}$.

\subsection{Capacity and cutoff rate analysis}\label{Subsect:SCACapacityCutoff}
For capacity and cutoff rate analysis of the SC architecture, we need to first specify a probabilistic model that covers all parts of the system.
As usual, we will use upper-case notation to denote the random variables and vectors in the system. 
In particular, we will write $X^N$ to denote the random vector at the output of the 1-1 mapper $f_N$; likewise, we will write $Y^N$ to denote the random vector at the input of the SDG $g_N$.
We will assume that $X^N$ is uniformly distributed,
$$
p_{X^N}(x^N) = 1/2^N, \quad \text{for all $x^N\in \{0,1\}^N$}.
$$
Since $X^N$ and $Y^N$ are connected by $N$ independent copies of $W$, we will have
$$
p_{Y^N|X^N}(y^N|x^N) =\prod_{i=1}^N W(y_i|x_i).
$$
The ensemble $(X^N,Y^N)$, thus specified, will serve as the core of the probabilistic analysis.
Next, we expand the probabilistic model to cover other signals of interest in the system.
We define $U^N$ as the random vector that appears at the output of the CEs in Fig.~\ref{fig:DerivedChannel2}.
Since $U^N$ is in 1-1 correspondence with $X^N$, it is uniformly distributed.
We define $\hat{U}^N$ as the random vector that the SDs feed back to the SDG as the estimate of $U^N$.
Ordinarily, any practical system has some non-zero probability that $\hat{U}^N\neq U^N$.
However, modeling such decision errors and dealing with the consequent error propagation effects in the SC chain is a difficult problem.
To avoid such difficulties, we will assume that the outer code in Fig.~\ref{fig:DerivedChannel2} is perfect, so that
\begin{equation}\label{eq:ModelAssumption}
\Pr(\hat{U}^N=U^N) = 1.
\end{equation}
This assumption eliminates the complications arising from error-propagation; however, the capacity and cutoff rates calculated under this assumption 
will be optimistic estimates of what can be achieved by any real system.
Still, the analysis will serve as a roadmap and provide benchmarks for practical system design.
(In the case of polar codes, we will see that the estimates obtained under the above ideal system model are in fact achievable.)
Finally, we define the soft-decision random vector $L^N$ at the output of the SDG so that 
its $i$th coordinate is given by
$$
L_i \defn (Y^N,U^{i-1}), \quad 1\le i\le N.
$$
(If it were not for the modeling assumption \eqref{eq:ModelAssumption}, it would be appropriate to use $\hat{U}^{i-1}$
in the definition of $L_i$ instead of $U^{i-1}$.)
This completes the specification of the probabilistic model for all parts of the system. 

We first focus on the capacity of the channel $W_N$ created by combining $N$ copies of $W$.
Since we have specified a uniform distribution for channel inputs,
the applicable notion of capacity in this analysis is the {\sl symmetric} capacity, defined as
\begin{equation*}
C_\sym(W) \defn C(W,\Qsym),
\end{equation*}
where $\Qsym$ is the uniform distribution, $\Qsym(0)=\Qsym(1)=1/2$.
Likewise, the applicable cutoff rate is now the symmetric one, defined as
\begin{equation*}
R_{0,\sym}(W) \defn R_0(W,\Qsym).
\end{equation*}

In general, the symmetric capacity may be strictly less than the true capacity, so there is a penalty for using a uniform distribution at channel inputs.
However, since we will be dealing with linear codes, the uniform distribution is the only appropriate distribution here.
Fortunately, for many channels of practical interest, the uniform distribution is actually the optimal one for 
achieving the channel capacity and the cutoff rate. These are the class of symmetric channels.
A binary-input channel is called {\sl symmetric} if for each output letter $y$ there exists a ``paired'' output letter $y'$ (not necessarily distinct from $y$) such that $W(y|0) = W(y'|1)$. 
Examples of symmetric channels include the BSC, the BEC, and the additive Gaussian noise channel with binary inputs.
As shown in \cite[p.~94]{GallagerBook}, for a symmetric channel, the symmetric versions of channel capacity and cutoff rate coincide with the true ones. 

We now turn to the analysis of the capacities of the bit-channels created by the SC architecture.
The SC architecture splits the vector channel $W_N$ into $N$ bit-channels, which we will denote by $W_N^{(i)}$, $1\le i\le N$.
The $i$th bit-channel connects the output $U_i$ of CE$_i$ to the input $L_i$ of SD$_i$,
$$
W_N^{(i)}:U_i \to L_i=(Y^N,U^{i-1}),
$$
and has symmetric capacity
\begin{equation*}
C_{\sym}(W_N^{(i)}) = I(U_i;L_i) = I(U_i;Y^N,U^{i-1}).
\end{equation*}
Here, $I(U_i;L_i)$ denotes the mutual information between $U_i$ and $L_i$. In the following analysis, we will be using the mutual information function and some of its basic properties, such as the chain rule. We refer to \cite[Ch.~2]{CoverThomas} for definitions and a discussion of such basic material.

The aggregate symmetric capacity of the bit-channels is calculated as
\begin{align}
\sum_{i=1}^N & \;C_\sym(W_N^{(i)}) = \sum_{i=1}^N  \;I(U_i;Y^NU^{i-1}) \nonumber\\
& \stackrel{(1)}{=}\sum_{i=1}^N  I(U_i;Y^N|U^{i-1}) \stackrel{(2)}{=} I(U^N;Y^N)\nonumber\\
& \stackrel{(3)}{=} I(X^N;Y^N) \stackrel{(4)}{=}\sum_{i=1}^N I(X_i;Y_i) = NC_{\text{sym}}(W) \label{eq:CapConGen}
\end{align}
where the equality (1) is due to the fact that $U_i$ and $U^{i-1}$ are independent, (2) is by the chain rule, (3) by the 1-1 property of $f_N$,
(4) by the memoryless property of the channel $W$.
Thus, the aggregate symmetric capacity of the underlying $N$ copies of $W$ is preserved by the combining and splitting operations.

Our main interest in using the SC architecture is to obtain a gain in the aggregate cutoff rate. 
We define the normalized symmetric cutoff rate under SC decoding as
\begin{equation}\label{eq:R0Ave}
\overline{R}_{0,\sym}(W_N) \defn \frac{1}{N}\sum_{i=1}^N R_{0,\sym}(W_N^{(i)}).
\end{equation}
The objective in applying the SC architecture may be stated as devising schemes for which 
\begin{align}\label{eq:CutoffGain}
\overline{R}_{0,\sym}(W_N) > R_{0,\text{sym}}(W)
\end{align}
holds by a significant margin.
The ultimate goal would be to have $\overline{R}_{0,\sym}(W_N)$ approach $C_\sym(W)$ as $N$ increases.

For specific examples of schemes that follow the SC architecture and achieve cutoff rate gains in the sense of \eqref{eq:CutoffGain},
we refer to \cite{Arikan2006} and the references therein.
We must mention in this connection that the multilevel coding scheme of Imai and Hirakawa \cite{imai1977new} is perhaps the first example of the SC architecture in the literature,
but the focus there was not to boost the cutoff rate.
In the next section, we discuss polar coding as another scheme that conforms to the SC architecture and provides the type of cutoff rate gains 
envisaged by \eqref{eq:CutoffGain}.

\section{Polar coding}\label{Sect:Polar}

Polar coding is an example of a coding scheme that fits into the framework of the preceding section
and has the property that 
\begin{equation}\label{eq:R0sumtoCap}
\overline{R}_{0,\sym}(W_N) \to C_\sym(W) \quad \text{as $N\to \infty$}.
\end{equation}
The name ``polar'' refers to a phenomenon called ``polarization'' that will be described later in this section.
We begin by describing the channel combining and splitting operations in polar coding.

\subsection{Channel combining and splitting for polar codes}\label{Sect:BasicConstruction}
The channel combining and splitting operations in polar coding follow the general principles already described in
detail in Sect.~\ref{Subsect:SCA}.
We only need to describe the particular 1-1 transformation $f_N$ that is used for constructing a polar code
of size $N$. We will begin this description starting with $N=2$.
 
\begin{figure}[ht]
\begin{center}
\hspace*{0cm}\resizebox{!}{1.2in}{
\includegraphics{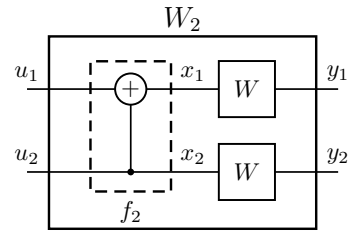}
}
\end{center}
\caption{Basic polar code construction.}
\label{fig:Polar0}
\end{figure}

The basic module of the channel combining operation in polar coding is shown in Fig.~\ref{fig:Polar0},
in which two independent copies of $W:\{0,1\}\to \cY$
are combined into a channel $W_2:\{0,1\}^2 \to \cY^2$ using 
a 1-1 mapping $f_2$ defined by
\begin{equation}\label{eq:BasicTransform}
f_2(u_1,u_2) \defn (u_1\oplus u_2,u_2)  
\end{equation}
where $\oplus$ denotes modulo-2 addition in the binary field $\F_2=\{0,1\}$. 
We call the basic transform $f_2$ the {\sl kernel} of the construction.
(We defer the discussion of how to find a suitable kernel until the end of this subsection.)

\begin{figure}[ht]
\begin{center}
\hspace*{0cm}\resizebox{!}{2.4in}{
\includegraphics{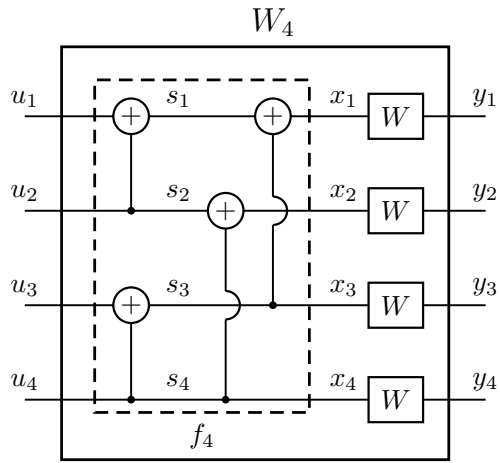}
}
\end{center}
\caption{Size-4 polar code construction.}
\label{fig:Polar2}
\end{figure}

Polar coding extends the above basic combining operation recursively to constructions of size $N=2^n$, for any $n\ge1$.
For $N=4$, the polar code construction is shown in Fig.~\ref{fig:Polar2},
where 4 independent copies of $W$ are combined by a 1-1 mapping $f_4$ into a channel $W_4$.

\begin{figure}[ht]
\begin{center}
\hspace*{-0cm}\resizebox{!}{1.6in}{
\includegraphics{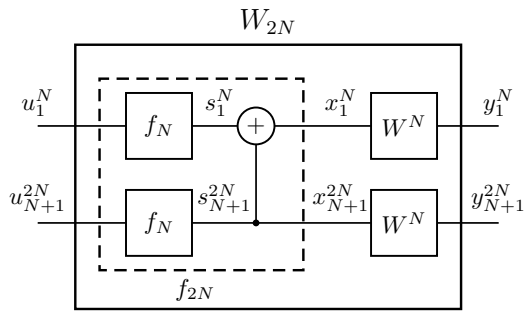}
}
\end{center}
\caption{Recursive extension of the polar code construction.}
\label{fig:Polar3}
\end{figure}

The general form of the recursion in polar code construction is illustrated in Fig.~\ref{fig:Polar3}
and can be expressed algebraically as
\begin{equation}\label{eq:deff2N}
f_{2N}(u^{2N}) \defn (f_N(u^N)\oplus f_N(u_{N+1}^{2N}),f_N(u_{N+1}^{2N})),
\end{equation}
where $\oplus$ denotes the componentwise mod-2 addition of two vectors of the same length over $\F_2$.

The transform $x^N=f_{N}(u^N)$ is {\sl linear} over the vector space $(\F_2)^N$, and can be expressed as
$$
x^N = u^N F_N,
$$
where $u^N$ and $x^N$ are row vectors, and $F_N$ is a matrix
defined recursively as 
$$
F_{2N} = \begin{bmatrix} F_N & 0_N \\ F_N & F_N\end{bmatrix},
\quad \text{with $F_2\defn \begin{bmatrix} 1 & 0 \\ 1 & 1\end{bmatrix}$}, 
$$
or, simply as
$$
F_N = (F_2)^{\otimes n}, \quad n = \log N,
$$
where the ``$\otimes$'' in the exponent denotes Kronecker power of a matrix \cite{ArikanIT2009}.

The recursive nature of the mapping $f_{N}$ makes it possible to compute 
$f_N(u^N)$ in time complexity $O(N\log N)$.
The polar transform $f_N$ is a ``fast'' transform over the field $\F_2$, akin to the ``fast Fourier transform'' of signal processing.

We wish to comment briefly on how to select a suitable kernel (basic module such as $f_2$ above) to get the polar code construction started.
In general, not only the kernel, but its size is also a design choice in polar coding; however, all else being equal, it is advantageous to use a small kernel to keep the complexity low. 
The specific kernel $f_2$ above has been found by exhaustively studying all $4!$ alternatives for a kernel of size $N=2$, corresponding to all permutations (1-1 mappings) of binary vectors
$(u_1,u_2)\in \{0,1\}^2$. 
Six of the permutations are listed in Table~\ref{table:permutations}.

\begin{table}[!h]
\renewcommand{\arraystretch}{1.3}
\caption{Basic permutations of $(u_1,u_2)$}
\centering
\begin{tabular}{c|c|c|c|c|c}
\hline
No. & 00 & 01 & 10 & 11 & Matrix\\\hline\hline
1 & 00 & 01 & 10 & 11 &  $\left[\begin{smallmatrix} 1 & 0 \\ 0 & 1\end{smallmatrix} \right]$\\\hline
2 & 00 & 10 & 01 & 11 &  $\left[\begin{smallmatrix} 0 & 1 \\ 1 & 0\end{smallmatrix} \right]$\\\hline
3 & 00 & 11 & 10 & 01 &  $\left[\begin{smallmatrix} 1 & 0 \\ 1 & 1\end{smallmatrix} \right]$\\\hline
4 & 00 & 01 & 11 & 10 &  $\left[\begin{smallmatrix} 1 & 1 \\ 0 & 1\end{smallmatrix} \right]$\\\hline
5 & 00 & 11 & 01 & 10 &  $\left[\begin{smallmatrix} 0 & 1 \\ 1 & 1\end{smallmatrix} \right]$\\\hline
6 & 00 & 10 & 11 & 01 &  $\left[\begin{smallmatrix} 1 & 1 \\ 1 & 0\end{smallmatrix} \right]$\\\hline
\end{tabular}
\renewcommand{\arraystretch}{1}
\label{table:permutations}
\end{table}

The title row displays the regular order of elements in $\{0,1\}^2$, each subsequent row displays a particular permutation of the same elements.
Each permutation listed in the table happens to be a linear transformation of $(u_1,u_2)$, with a transformation matrix as shown as the final entry of the related row.
The remaining 18 permutations that are not listed in the table can be obtained as affine transformations of the six that are listed.
For example, by adding (mod-2) a non-zero constant offset vector, such as $10$, to each entry in the table, we obtain six additional permutations.

The first and the second permutations in the table are trivial permutations that provide no channel combining.
The third and the fifth permutations are equivalent from a coding point of view; their matrices are column permutations of each other,
which corresponds to permuting the elements of the codeword during transmission---an operation that has no effect on the capacity or cutoff rate.
The fourth and the sixth permutations are also equivalent to each other in the same sense that the third and the fifth permutations are.
The fourth permutation is not suitable for our purposes since it does not provide any channel combining (entanglement) under the decoding order $u_1$ first, $u_2$ second.
For the same reason, the sixth permutation is not suitable, either.
The third and the fifth permutations (and their affine versions) remain as the only viable alternatives; and they are all equivalent from a capacity/cutoff rate viewpoint.
Here, we use the third permutation since it is the simplest one among the eight viable candidates.

\subsection{Capacity and cutoff rate analysis for polar codes}\label{Sect:AnalysisPolar}
For the analysis in this section, we will use the general setting and notation of Sect.~\ref{Subsect:SCACapacityCutoff}.
We begin our analysis with the case $N=2$.
The basic transform \eqref{eq:BasicTransform} creates a channel $W_2: (U_1,U_2)\to (Y_1,Y_2)$
with transition probabilities
\begin{equation*}
W_2(y_1,y_2|u_1,u_2) = W(y_1|u_1\oplus u_2)W(y_2|u_2).
\end{equation*}
This channel is split by the SC scheme into two bit-channels
$W_2^{(1)}:U_1\to (Y_1,Y_2)$ and $W_2^{(2)}: U_2\to (Y_1,Y_2,U_1)$
with transition probabilities
\begin{gather*}
W_2^{(1)}(y_1y_2|u_1)=\sum_{u_2\in \{0,1\}} \Qsym(u_2) W(y_1|u_1\oplus u_2)W(y_2|u_2),\\
W_2^{(2)}(y_1y_2u_1|u_2)  = \Qsym(u_1) W(y_1|u_1\oplus u_2)W(y_2|u_2).
\end{gather*}

Here, we introduce the alternative notation $W^-$ and $W^+$ to denote $W_2^{(1)}$ and $W_2^{(2)}$, respectively.
This notation will be particularly useful in the following discussion.
We observe that the channel $W^-$ treats $U_2$ as pure noise; while, $W^+$ treats $U_1$ as an observed (known) entity.
In other words, the transmission of $U_1$ is hampered by interference from $U_2$; while $U_2$ ``sees'' a channel of diversity order two,
after ``canceling'' $U_1$.
Based on this interpetation, we may say that the polar transform creates a ``bad'' channel $W^-$ and a ``good'' channel $W^+$.
This statement can be justified by looking at the capacities of the two channels.
 
The symmetric capacities of $W^-$ and $W^+$ are given by
\begin{equation*}
C_\sym(W^-) = I(U_1;Y_1Y_2),\quad C_\sym(W^+)= I(U_2;Y_1Y_2U_1).
\end{equation*}
We observe that
\begin{equation}\label{eq:CapCon}
C_\sym(W^-) + C_\sym(W^+) = 2C_\sym(W),
\end{equation}
which is a special instance of the general conservation law \eqref{eq:CapConGen}.
The symmetric capacity is conserved, but redistributed unevenly. It follows from
basic properties of mutual information function that
\begin{equation}\label{eq:CapIneq}
C_\sym(W^-) \,\le\, C_\sym(W) \,\le\, C_\sym(W^+),
\end{equation}
where the inequalities are strict unless $C_\sym(W)$ equals 0 or 1.
For a proof, we refer to \cite{ArikanIT2009}.

We will call a channel $W$ {\sl extreme} if $C_\sym(W)$ equals 0 or 1.
Extreme channels are those for which there is no need for coding: if $C_\sym(W)=1$, one can send data uncoded;
if $C_\sym(W)=0$, no code will help. 
Inequality \eqref{eq:CapIneq} states that, unless the channel $W$ is extreme, the size-2 polar transform creates a
channel $W^+$ that is strictly better than $W$, and a second channel $W^-$ that is strictly worse than $W$.
By doing so, the size-2 transform starts the polarization process.

As regards the cutoff rates, we have
\begin{equation}\label{eq:R0gain}
R_{0,\sym}(W^-) + R_{0,\sym}(W^+) \ge 2 R_{0,\sym}(W),
\end{equation}
where the inequality is strict unless $W$ is extreme.
This result, proved in \cite{ArikanIT2009}, states that the basic transform 
{\sl always} creates a cutoff rate gain, except when $W$ is extreme. 

An equivalent form of \eqref{eq:R0gain}, which is the one that was actually proved in \cite{ArikanIT2009}, is the following inequality about the Bhattacharyya parameters,
\begin{equation}\label{eq:Zloss}
Z(W^-) + Z(W^+) \,\le\, 2\,Z(W),
\end{equation}
where strict inequality holds unless $W$ is extreme.
The equivalence of \eqref{eq:R0gain} and \eqref{eq:Zloss} is easy to see from the
relation
\begin{equation*}
R_{0,\sym}(W) = 1 -\log[1 + Z(W)],
\end{equation*}
which is a special form of \eqref{eq:defR0BinaryQ} with $Q=\Qsym$.

Given that the size-2 transform improves the cutoff rate of any given channel $W$ (unless $W$ is already extreme in which case there is no need to do anything), it is natural to seek 
methods of applying the same method recursively so as to gain further improvements.
This is the main intuitive idea behind polar coding.
As the cutoff-rate gains are accummulated over each step of recursion, the synthetic bit-channels that are created in the process
keep moving towards extremes.

To see how recursion helps improve the cutoff rate as the size of the polar transform is doubled, let us consider the next step of the construction, $N=4$ .
The key recursive relationships that tie the size-4 construction to size-2 construction are the following:
\begin{align} 
W_4^{(1)} \equiv (W_2^{(1)})^-,\qquad &
W_4^{(2)} \equiv (W_2^{(1)})^+,\label{eq:firstequivalence}\\
W_4^{(3)} \equiv (W_2^{(2)})^-,\qquad &
W_4^{(4)} \equiv (W_2^{(2)})^+.\label{eq:secondequivalence}
\end{align}
The first claim $W_4^{(1)} \equiv (W_2^{(1)})^-$ means that $W_4^{(1)}$ is equivalent to the bad channel obtained by applying a size-2 transform on two independent copies of $W_2^{(1)}$. 
The other three claims can be interpreted similarly.

To prove the validity of \eqref{eq:firstequivalence} and \eqref{eq:secondequivalence}, let us refer to 
Fig.~\ref{fig:Polar2} again.
Let $(U^4,S^4,X^4,Y^4)$ denote the ensemble of random vectors that correspond to the signals $(u^4,s^4,x^4,y^4)$ 
in the polar transform circuit.
In accordance with the modeling assumptions of Sect.~\ref{Subsect:SCACapacityCutoff}, the random vector $X^4$ is uniformly distributed over $\{0,1\}^4$. 
Since both $S_1^4$ and $U^4$ are in 1-1 correspondence with $X_1^4$, they, too, are uniformly distributed over $\{0,1\}^4$.
Furthermore, the elements of $X^4$ are i.i.d. uniform over $\{0,1\}$, and similarly for the elements of $S^4$, and of $U^4$.

\begin{figure}[ht]
\begin{center}
\hspace*{0cm}\resizebox{!}{1.6in}{
\includegraphics{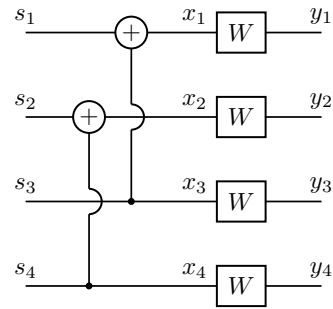}
}
\end{center}
\caption{Intermediate stage of splitting for size-4 polar code construction.}
\label{fig:Polar5}
\end{figure}

Let us now focus on Fig.~\ref{fig:Polar5} which depicts the relevant part of Fig.~\ref{fig:Polar2} for the present discussion.
Consider the two channels
$$
W':S_1\to (Y_1,Y_3), \quad W'': S_2\to (Y_2,Y_4),
$$
embedded in the diagram.
It is clear that
$$
W' \equiv W'' \equiv W_2^{(1)} \equiv W^-.
$$
Furthermore, the two channels $W'$ and $W''$ are independent.
This is seen by noticing that $W'$ is governed by the set of random variables $(S_1,S_3,Y_1,Y_3)$, which is disjoint from the set of variables
$(S_2,S_4,Y_2,Y_4)$ that govern $W''$. 

\begin{figure}[ht]
\begin{center}
\hspace*{0cm}\resizebox{!}{0.8in}{
\includegraphics{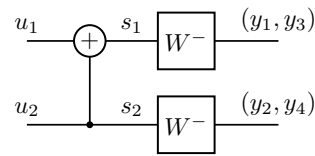}
}
\end{center}
\caption{A size-2 polar code construction embedded in a size-4 construction.}
\label{fig:Polar6}
\end{figure}

Returning to the size-4 construction of Fig.~\ref{fig:Polar2}, we now see that the effective channel seen by the pair of inputs 
$(U_1,U_2)$ is the combination of $W'$ and $W''$, or equivalently, of two independent copies of $W_2^{(1)}\equiv W^-$, as shown in Fig.~\ref{fig:Polar6}.
The first pair of claims \eqref{eq:firstequivalence} follows immediately from this figure.

\begin{figure}[ht]
\begin{center}
\hspace*{0cm}\resizebox{!}{0.8in}{
\includegraphics{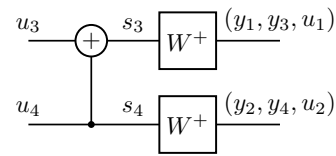}
}
\end{center}
\caption{A second size-2 polar code construction inside a size-4 construction.}
\label{fig:Polar7}
\end{figure}

The second pair of claims \eqref{eq:secondequivalence} follows by observing that, after decoding $(U_1,U_2)$, 
the effective channel seen by the pair of inputs $(U_3,U_4)$ is the combination of two independent copies of $W_2^{(2)}\equiv W^+$ as shown in Fig.~\ref{fig:Polar7}. 

The following conservation rules are immediate from \eqref{eq:CapCon}.
\begin{gather*}
C_\sym(W^{--}) + C_\sym(W^{-+}) \,= \, 2\,C_\sym(W^-),\\
C_\sym(W^{+-}) + C_\sym(W^{++}) \,= \, 2\,C_\sym(W^{+}).
\end{gather*}
Likewise, we have, from \eqref{eq:R0gain},
\begin{gather*}
R_{0,\sym}(W^{--}) + R_{0,\sym}(W^{-+}) \,\ge \, 2\,R_{0,\sym}(W^{-}),\\
R_{0,\sym}(W^{+-}) + R_{0,\sym}(W^{++}) \, \ge\, 2\,R_{0,\sym}(W^{+}).
\end{gather*}
Here, we extended the notation and used $W^{--}$ to denote $(W^-)^-$, and similarly for
$W^{-+}$, etc.

If we normalize the aggregate cutoff rates for $N=4$ and compare with the normalized cutoff rate for $N=2$,
we obtain 
\begin{gather*}
\overline{R}_{0,\sym}(W_4) \,\ge\,  
\overline{R}_{0,\sym}(W_2) \,\ge\,  
R_{0,\sym}(W).
\end{gather*}
These inequalities are strict unless $W$ is extreme.
 
The recursive argument given above can be applied to the situtation in
Fig.~\ref{fig:Polar3} to show that for any $N=2^n$, $n\ge 1$,
and $1\le i\le N$, the following relations hold
\begin{equation*}
W_{2N}^{(2i-1)} \equiv (W_N^{(i)})^-, \quad W_{2N}^{(2i)} \equiv (W_N^{(i)})^+,
\end{equation*}
from which it follows that
\begin{gather*}
\overline{R}_{0,\sym}(W_{2N}) \,\ge\,  
\overline{R}_{0,\sym}(W_N).
\end{gather*}

These results establish that the sequence of normalized cutoff rates $\{\overline{R}_{0,\sym}(W_N)\}$ is monotone non-decreasing in $N$.
Since $\overline{R}_{0,\sym}(W_N) \le C_\sym(W)$ for all $N$, the sequence must converge to 
a limit.
It turns out, as might be expected, that this limit is the symmetric capacity $C_\sym(W)$.
We examine the asymptotic behavior of the polar code construction process in the next subsection.

\subsection{Polarization and elimination of the outer code}

As the construction size in the polar transform is increased, gradually a ``polarization'' phenomenon takes holds.
All channels $\{W_N^{(i)}:1\le i\le N\}$ created by the polar transform, except for a vanishing fraction, approach extreme limits (becoming near perfect or useless)
with increasing $N$.
One form of expressing polarization more precisely is the following. For any fixed $\delta >0$, the channels created by the polar transform satisfy
\begin{equation}\label{eq:polarization1}
\frac1N \left|\bigl\{i: R_{0,\sym}(W_N^{(i)}) > 1-\delta\bigr\}\right| \to C_\sym(W)
\end{equation}
and 
\begin{equation}\label{eq:polarization2}
\frac1N \left|\bigl\{i: R_{0,\sym}(W_N^{(i)}) < \delta\bigr\}\right| \to 1-C_\sym(W)
\end{equation}
as $N$ increases. 
($|\cA|$ denotes the number of elements in set $\cA$.)
A proof of this result using martingale theory can be found in \cite{ArikanIT2009};
for a recent simpler proof that avoids martingales, we refer to \cite{alsan2014simple}.

As an immediate corollary to \eqref{eq:polarization1}, we obtain \eqref{eq:R0sumtoCap}, establishing the main goal of this analysis.
While this result is very reassuring, there are many remaining technical details that have to be taken care of
before we can claim to have a practical coding scheme.
First of all, we should not forget that the validity of \eqref{eq:R0sumtoCap} rests on the assumption \eqref{eq:ModelAssumption} that there are no errors 
in the SC decoding chain.
We may argue that we can satisfy assumption \eqref{eq:ModelAssumption} to any desired degree of accuracy by using convolutional codes of sufficiently long constraint lengths.
Luckily, it turns out using such convolutional codes is unnecessary to have a practically viable scheme.
The polarization phenomenon creates sufficient number of sufficiently good channels fast enough that the validity of \eqref{eq:R0sumtoCap} can be maintained without 
any help from an outer convolutional code and sequential decoder.
The details of this last step of polar code construction are as follows.

Let us reconsider the scheme in  Fig.~\ref{fig:DerivedChannel2}.
At the outset, the plan was to operate the $i$th convolutional encoder CE$_i$ at a rate just below the symmetric cutoff rate $R_{0,\sym}(W_N^{(i)})$. 
However, in light of the polarization phenomenon, we know that almost all the cutoff rates $R_{0,\sym}(W_N^{(i)})$ are clustered around 0 or 1 for $N$ large.
This suggests rounding off the rates of all convolutional encoders to 0 or 1, effectively eliminating the outer code.
Such a revised scheme is highly attractive due to its simplicity, but dispensing with the outer code exposes the system to
unmitigated error propagation in the SC chain. 

To analyze the performance of the scheme that has no protection by an outer code, let $\cA$ denote the set of indices $i\in \{1,\ldots,N\}$ of input variables
$U_i$ that will carry data at rate 1. We call $\cA$ the set of ``active'' variables. 
Let $\cA^c$ denote the complement of $\cA$, and call this set the set of ``frozen'' variables.
We will denote the active variables collectively by $U_{\cA}\defn (U_i:i\in\cA)$ and the frozen ones by $U_{\cA^c}\defn (U_i:i\in cA^c)$,
each vector regarded as a subvector of $U^N$. Let $K$ denote the size of $\cA$.

Encoding is done by setting
$U_{\cA}=D^K$ and $U_{\cA^c}=b^{N-K}$ where $D^K$ is user data equally likely to take any value in $\{0,1\}^K$ and
$b^{N-K}\in \{0,1\}^{N-K}$ is a fixed pattern.
The user data may change from block to the next, but the frozen pattern remains the same and is known to the decoder.
This system carries $K$ bits of data in each block of $N$ channel uses, for a transmission rate of $R=K/N$.

At the receiver, we suppose that there is an SC decoder that computes its decision $\hat{U}^N$ by calculating the likelihood ratio
$$
L_i \defn \Pr(U_i = 0|Y^N,\hat{U}^{i-1})\big/\Pr(U_i = 1|Y^N,\hat{U}^{i-1}),
$$
and setting 
$$
\hat{U}_i =\begin{cases} U_i, & \text{if $i\in \cA^c$},\\
0, & \text{if $i\in \cA$ and $L_i >1$};\\
1, & \text{if $i\in \cA$ and $L_i \le 1$},
\end{cases}
$$
successively, starting with $i=1$.
Since the variables $U_{\cA^c}$ are fixed to $b^{N-K}$, this decoding rule can be implemented at the decoder.
The probability of frame error for this system is given by
\begin{equation*}
P_e(\cA,b^{N-K}) \defn P(\hat{U}_{\cA}\neq U_{\cA}|U_{\cA^c}=b^{N-K}).
\end{equation*}

For a symmetric channel, the error probability $P_e(\cA,b^{N-K})$ does not depend on $b^{N-K}$ \cite{ArikanIT2009}.
A convenient choice in that case may be to set $b^{N-K}$ to the zero vector. 
For a general channel, we consider the average of the error probability over all possible choices for $b^{N-K}$,
namely,
\begin{equation*}
P_e(\cA) \defn \frac{1}{2^{(N-K)}} \sum_{b^{N-K}\in \{0,1\}^{N-K}} P(\hat{U}_{\cA}\neq U_{\cA}|U_{\cA^c}=b^{N-K}).
\end{equation*}
It is shown in \cite{ArikanIT2009} that
\begin{equation}\label{eq:PeBoundPolar}
P_e(\cA) \le \sum_{i\in \cA} Z(W_N^{(i)}),
\end{equation}
where $Z(W_N^{(i)})$ is the Bhattacharyya parameter of $W_N^{(i)}$.

The bound \eqref{eq:PeBoundPolar} suggests that $\cA$ should be chosen so as to minimize the upper bound on $P_e(\cA)$.
The performance attainable by such a design rule can be calculated directly from 
the following polarization result from \cite{ArikanTelatarISIT2009}.

For any fixed $\beta <\frac12$, the Bhattacharyya parameters created by the polar transform satisfy
\begin{equation}\label{eq:polarization3}
\frac1N \left|\bigl\{i: Z(W_N^{(i)}) < 2^{-N^{\beta}}\bigr\}\right| \to C_\sym(W).
\end{equation}

In particular, if we fix $\beta =0.49$, the fraction of channels 
$W_N^{(i)}$ in the population $\{W_N^{(i)}:1\le i\le N\}$
satisfying
\begin{equation}\label{eq:condZ}
Z(W_N^{(i)})< 2^{-N^{0.49}}
\end{equation}
approaches the symmetric capacity $C_\sym(W)$ as $N$ becomes large. 
So, if we fix the rate $R < C_\sym(W)$, then for all $N$ sufficiently large,
we will be able to select an active set $\cA_N$ of size $K=NR$ such that
\eqref{eq:condZ} holds for each $i\in \cA_N$.
With $\cA_N$ selected in this way, the probability of error is bounded as
\begin{equation*}
P_e(\cA_N)  < \sum_{i\in \cA_N} Z(W_N^{(i)}) \le N2^{-N^{0.49}} \to 0.
\end{equation*}
This establishes the feasibility of constructing polar codes that operate at any rate $R <C_\sym(W)$ with a probability of error going to 0 exponentially as $\approx 2^{-\sqrt{N}}$.

This brings us to the end of our discussion of polar codes.
We will close by mentioning two important facts that relate to complexity. The SC decoding algorithm for polar codes can be implemented in complexity $O(N\log N)$
\cite{ArikanIT2009}. 
The contruction complexity of polar codes, namely, the selection of an optimal (subject to numerical precision) set $\cA$ of active channels (either by computing the Bhattacharyya parameters $\{Z_N^{(i)}\}$
or some related set of quality parameters) can be done in complexity $O(N)$, as shown in the sequence of papers \cite{mori2009performance}, \cite{tal2013construct}, and \cite{pedarsani2011construction}.

\section{Summary}\label{Sect:Conclusion}

In this paper we gave an account of polar coding from a historical perspective, tracing the original line of thinking that led to its development.

The key motivation for polar coding was to boost the cutoff rate of sequential decoding.
The schemes of Pinsker and Massey suggested a two-step mechanism: first build a vector channel from independent copies of a given channel; next, split the vector channel into correlated subchannels. 
With proper combining and splitting, it is possible to obtain an improvement in the aggregate cutoff rate.
Polar coding is a recursive implementation of this basic idea. 
The recursiveness renders polar codes both analytically tractable, which leads to an explicit code construction algorithm, and also makes it possible to
encode and decode these codes at low-complexity. 

Although polar coding was originally intended to be the inner code in a concatenated scheme, it turned out (to our pleasant surprise) that the inner code was so reliable that there was no need for the outer convolutional code or the sequential decoder. However, to further improve polar coding, one could still consider adding an outer coding scheme, as originally planned.

\appendix[Derivation of the pairwise error bound]

This appendix provides a proof of the pairwise error bound \eqref{eq:pairwiseerrorbound}.
The proof below is standard textbook material. It is reproduced here for completeness and to demonstrate the simplicity of the basic idea underlying the cutoff rate.

Let $\cC=\{x^N(1),\ldots,x^N(M)\}$ be a specific code, let $R=(1/N) \log M$ be the rate of the code.
Fix two distinct messages $m\neq m'$, $1\le m,m'\le M$, and define $P_{m,m'}(\cC)$ as in \eqref{eq:defPmm'}.
Then, 
\begin{align*}
P_{m,m'}(\cC) &\defn \sum_{y^N\in E_{m,m'}(\cC)} W^N(y^N|x^N(m))\\
& \stackrel{(1)}{\le} \sum_{y^N}  W^N(y^N|x^N(m))\sqrt{\frac{W^N(y^N|x^N(m'))}{W^N(y^N|x^N(m))}} \\
& = \sum_{y^N}  \sqrt{W^N(y^N|x^N(m))W^N(y^N|x^N(m'))}\\
& \stackrel{(2)}{=} \prod_{n=1}^N \left[\sum_{y_n}  \sqrt{W(y_n|x_n(m))W(y_n|x_n(m'))}\right]\\
& \stackrel{(3)}{=} \prod_{n=1}^N Z_{m,m'}(n),
\end{align*}
where the inequality (1) follows by the simple observation that
$$
\sqrt{\frac{W^N(y^N|x^N(m'))}{W^N(y^N|x^N(m))}} \ge \begin{cases} 1, & y^N\in E_{m,m'}(\cC)\\
0, & y^N\notin E_{m,m'}(\cC).
\end{cases},
$$
equality (2) follows by the memoryless channel assumption,
and (3) by the definition
\begin{equation}\label{eq:defZmm}
Z_{m,m'}(n)\defn \sum_{y} \sqrt{W(y|x_n(m))W(y|x_n(m'))}.
\end{equation}
At this point the analysis becomes dependent on the specific code structure.
To continue, we consider the ensemble average of the pairwise error probability, $\overline{P}_{m,m'}(N,Q)$, defined by \eqref{eq:defaveragePmm'}.
\begin{align*}
\overline{P}_{m,m'}(N,Q)  & \le \overline{\prod_{n=1}^N Z_{m,m'}(n)} \\
&  \stackrel{(1)}{=} \prod_{n=1}^N \overline{Z}_{m,m'}(n) \stackrel{(2)}{=} \left[\,\overline{Z}_{m,m'}(1)\right]^N 
\end{align*}
where the overbar denotes averaging with respect to the code ensemble, the equality (1) is due to the independence of the random variables $\{Z_{m,m'}(n):1\le n\le N\}$, and (2) is by the fact that the same set of random variables are identically-distributed.
Finally, we note that, 
\begin{align*}
\overline{Z}_{m,m'}(1) & = \overline{\sum_{y}\sqrt{W(y|X_1(m)) W(y|X_{1}(m'))}}  \nonumber\\
 & = \sum_{y} \bigg[\overline{\sqrt{W(y|X_1(m))}}\bigg]\,\bigg[\overline{\sqrt{W(y|X_1(m'))}}\bigg]  \nonumber\\
 & = \sum_{y}\bigg[ \sum_{x} Q(x)\sqrt{W(y|x)}\bigg]^2.
\end{align*}
In view of the definition \eqref{eq:defR0Q}, this proves that $\overline{P}_{m,m'}(N,Q)\le 2^{-NR_0(Q)}$, for any $m\neq m'$.

\section*{Acknowledgement}
The author would like to thank Senior Editor M. Pursley and anonymous reviewers for extremely helpful suggestions to improve the paper. This work was supported by the FP7 Network of Excellence NEWCOM\# under grant agreement~318306.

\end{document}